\documentstyle[aps,prd,psfig]{revtex}
\begin{document}
\renewcommand{\thesection}{\arabic{section}}
\renewcommand{\thesubsection}{\arabic{subsection}}

\title{Impossibility of Spin Polarized States for Neutron Star / Proto-neutron 
Star Matter in $\beta$-Equilibrium Condition}

\author{
 Sutapa Ghosh$^{a)}${\thanks{E-Mail: sutapa@klyuniv.ernet.in}},
 Soma Mandal$^{a)}$ {\thanks{E-Mail: soma@klyuniv.ernet.in}}
 and Somenath Chakrabarty$^{a),b)}${\thanks{E-Mail: 
 somenath@klyuniv.ernet.in}} }
\address{
$^{a)}$Department of Physics, University of Kalyani, Kalyani 741 235,
India and
$^{b)}$Inter-University Centre for Astronomy and Astrophysics, Post Bag 4,
Ganeshkhind, Pune 411 007, India
}
\date{\today}
\maketitle

\begin{abstract}
It is shown explicitly that a ferromagnetic transition of neutron star (NS) /
proto-neutron star (PNS) matter in the $\beta$-equilibrium condition
with $\sigma-\omega-\rho$ exchange type of mean field approximation  can
actually occur if and only if the neutrinos remain trapped within the system,
and perhaps it is also necessary for the neutrinos / anti-neutrinos to carry
some finite non-zero mass. It is further shown that the electrons also play 
a significant role in this transition process. It is therefore, very much 
unlikely that such a transition of spin polarization can really take place 
in old neutron stars of very low temperature, whereas, the possibility of 
spontaneous ferromagnetic transition cannot be ruled out in a newly born PNS.
\end{abstract}

\noindent The origin of strong magnetic field of pulsars, which
are also believed to be the rotating neutron stars, is not exactly known 
till today \cite{R1}. There are so many models of which the idea of
flux conservation during collapse gives rise to a field strength of 
$\approx 10^{12}$G for some class of neutron stars.
This field is known as fossil
remnant from the progenitor star, with a typical magnetic field strength
of $\sim 10^4$G. The other alternative model for the generation of
magnetic field is by some kind of dynamo process or electric current flowing in
highly conducting material of the neutron stars after their formation
\cite{R1,R2,R3,R4,R5}. Now the 
studies on magnetic field evolution of neutron stars by population synthesis
codes show that the magnetic field decay time scale for typical neutron
stars (not the magnetar like exotic objects) is $\sim 10^7-10^9$yrs and the 
residual magnetic
field $\sim 10^8$G \cite{R6} (see also \cite{R7,R8,R9}). This is also the 
order of magnitude of magnetic 
field strength for the millisecond pulsars, which are believed to be 
re-generated and started 
their rotations with very high angular velocity after crossing the death 
line from pulsar graveyard by the accretion  of
matter \cite{R1}, i.e., such objects are very old (and also cold) 
neutron stars. If this residual magnetic field is assumed to be permanent and 
not generated by some dynamo process (the assumption is possibly true in the
case of very old neutron stars) within the star, one of the possibility
of such permanent magnetic field is a spontaneous ferromagnetic
transition by the spin alignment of permanent elementary magnetic dipoles 
of neutrons, protons etc.  Since it is generally believed that such a transition
may be the origin of magnetic field, in particular the remnant magnetic
field of millisecond pulsars which are assumed to be very old neutron
star, we have considered it to be an important issue. Of course, if a
spontaneous ferro-magnetic transition is not possible in neutron star
matter in $\beta$-equilibrium in presence of non-degenerate neutrinos,
in spite of that
we realized that this very old problem of spontaneous ferro-magnetic
transition has become extremely interesting by itself (also has some
academic importance). To the best of
our knowledge such studies have not been done before and this is the
first attempt to investigate this anomaly.

In this article we shall show that a spontaneous ferromagnetic transition 
is completely forbidden in a NS if
the matter is in $\beta$-equilibrium and neutrinos leave
the system immediately after their formation (i.e., non-degenerate). However,
such a transition is possible if the produced neutrinos remain trapped
within the neutron matter system (i.e. in PNS) for quite a long time at
the initial stage of evolution. Now the ferromagnetic materials are
paramagnetic but, because of interactions between atoms, show
drastically different behavior. Below the Curie temperature,
ferromagnetic substance show spontaneous magnetization, that is all the
magnetic moments in a microscopically large region called a domain are
aligned. The application of an external field tend to cause the domains
to change and the moments in different domains to line up together
leading to the saturation of the bulk magnetization. Removal of the
field leaves a considerable fraction of the moments still aligned,
giving a permanent magnetization. We further assume a $\sigma-\omega-\rho$
meson type of mean field interaction picture in the hadronic sector,
whereas, electrons and neutrons are assumed to be non-interacting.

In a NS / PNS the dynamic nature of chemical equilibrium among the
constituents is mainly dominated  by the following
weak interaction processes (URCA processes)
\begin{eqnarray}
n &\rightarrow & p+ e^- +\bar \nu_e \nonumber \\
p+ e^- & \rightarrow & n + \nu_e
\end{eqnarray}
Now in a typical neutron star, considering the spontaneously spin
polarization of hadronic matter, in which neutrinos leave the
system as soon as they are produced, the conditions for dynamical
$\beta$-equilibrium is given by
\begin{eqnarray}
\mu_{n\uparrow}& = &\mu_{p\uparrow}+\mu_{e\uparrow}\nonumber \\
& = &\mu_{p\uparrow}+\mu_{e\downarrow}\nonumber \\
& = &\mu_{p\downarrow}+\mu_{e\uparrow}
\end{eqnarray}
\begin{eqnarray}
\mu_{n\downarrow}& = &\mu_{p\downarrow}+\mu_{e\downarrow}\nonumber \\
& = &\mu_{p\downarrow}+\mu_{e\uparrow}\nonumber \\
& = &\mu_{p\uparrow}+\mu_{e\downarrow}
\end{eqnarray}
where up and down arrows indicate two directions of spin.
In this case, the neutrinos which leave the system without delay,
carry the spin signature
of the processes. Of course, the neutrinos (anti-neutrinos) are assumed to
be massless and left-handed (right-handed), i.e. they have definite
helicity states. Now the general expression for the chemical potential
for proton or neutron is given by
\begin{eqnarray}
\mu_p&=&\epsilon_f^p+g_\omega \omega_0 +\frac{1}{2}g_\rho
\rho_3^0\nonumber \\
\mu_n&=&\epsilon_f^n+g_\omega \omega_0 -\frac{1}{2}g_\rho
\rho_3^0
\end{eqnarray}
where the first terms in these equations are given by
$\epsilon_f^i=(k_f^{i^2}+m^{*^2})^{1/2}$, the neutron-proton effective
single particle energy, $m^*=m-g_\sigma \sigma_0$ is the hadronic
effective mass, $g_i$'s are the
coupling constants, and
\begin{equation}
g_\sigma \sigma_0 =\left ( \frac{g_\sigma}{m_\sigma}\right )^2
\sum_{i=n,p}\frac {g_im^*}{2\pi^2} \sum_{\tau=\uparrow}^{\downarrow}
\int_0^{k_{F_i}^{(\tau)}} \frac{k^2dk}{(k^2+m^{*^2})^{1/2}}
\end{equation} 
Since there are sum over both iso-spin and spin, the explicit spin dependence 
of hadronic effective mass will not there, however there is an
indirect spin dependence through the fermi momenta ($k_{F_i}^{(\tau)}$). 
The degeneracy parameter
$g_i=1$ for both $i=n$ and $p$.
Now only the iso-spin $3$-component of the
iso-vector meson field has finite mean value, i.e., the neutral $\rho$
meson ($\rho_3^\mu$). The Lorentz three-current part must vanish in the
static ground state of matter, just as the vector meson $\omega^\mu$,
leaving only $\rho_3^0$. Since we are interested in spontaneous
ferro-magnetic transition at low temperature (old neutron stars) in
absence of external magnetic field, we have not considered the
anomalous magnetic moments of the constituents. It is necessary to take
into account the effect of anomalous magnetic moments to study the
properties of nuclear matter in presence of strong magnetic field.
From the above  two sets of equations 
it is very easy to establish the following relations (since the spin
dependence in chemical potentials come from the anomalous magnetic
moments and is given by $s \delta_i B$, where $i=p$, $n$ or $e$,
$s=+1$ for spin up and $s=-1$ for spin down cases, $\delta_i$ is related
to the anomalous magnetic moment and finally $B$ is the
strength of magnetic field).
\begin{equation}
\mu_{n\uparrow}-\mu_{n\downarrow}=
\mu_{p\uparrow}-\mu_{p\downarrow}=
\mu_{e\uparrow}-\mu_{e\downarrow}=0
\end{equation}
which further means
\begin{equation}
\epsilon_f^{p\uparrow}= \epsilon_f^{p\downarrow};
~~\epsilon_f^{n\uparrow}= \epsilon_f^{n\downarrow};
~~\epsilon_f^{e\uparrow}= \epsilon_f^{e\downarrow}
\end{equation}
Hence it is extremely trivial to show from the definition of number density,
that each species with up and down spins are equal in
number, i.e. $n_{p\uparrow}= n_{p\downarrow}$, 
$n_{n\uparrow}= n_{n\downarrow}$,
and $n_{e\uparrow}= n_{e\downarrow}$. 
{\sl{This is absolutely independent of the conventional models generally used 
to obtain the equation of states of NS matter}}. 
Therefore, it is unlikely to 
have a spin polarized NS matter or a spontaneous ferromagnetic transition in 
NS, in which neutrinos flow out of the system as soon as they are created. 
Next we consider the case of a PNS, which is the very early stage of evolution 
of a typical NS, neutrinos remain trapped within the matter and it
requires a finite duration of time (a few ms) to make the system neutrino 
free. In the case of PNS, since
neutrinos are degenerate, their chemical potentials cannot be neglected
in the dynamic equilibrium conditions as mentioned above (eqns.(2) and
(3)). Therefore, the
modified form of these two sets of equations are given by
\begin{eqnarray}
\mu_{n\uparrow}& = &\mu_{p\uparrow}+\mu_{e\uparrow} - \mu_{\nu
\downarrow}\nonumber \\
& = &\mu_{p\uparrow}+\mu_{e\downarrow} - \mu_{\nu \uparrow}\nonumber \\
& = &\mu_{p\downarrow}+\mu_{e\uparrow} -\mu_{\nu \uparrow}
\end{eqnarray}
\begin{eqnarray}
\mu_{n\downarrow}& = &\mu_{p\downarrow}+\mu_{e\downarrow} - \mu_{\nu
\uparrow}\nonumber \\
& = &\mu_{p\downarrow}+\mu_{e\uparrow} - \mu_{\nu \downarrow}\nonumber \\
& = &\mu_{p\uparrow}+\mu_{e\downarrow} -\mu_{\nu \downarrow}
\end{eqnarray}
From these two sets of conditions, it is very easy to show
\begin{equation}
\mu_{n\uparrow}-\mu_{n\downarrow}=
\mu_{p\uparrow}-\mu_{p\downarrow}=
\mu_{e\uparrow}-\mu_{e\downarrow}=
\mu_{\nu \uparrow}-\mu_{\nu \downarrow}=S {\rm{~~~(say)}}
\end{equation}
where $\vert S \vert \neq 0$. This parameter determines the degree of
polarization of the system, i.e. it controls the ferromagnetic transition
of the system. In the case of unpolarized matter, $\vert S \vert \approx
0$, whereas for the spontaneously created ferromagnetic system, it is 
extremely large. From eqn.(10), it is also obvious that the magnitude of the 
factor $S$ also significantly depends on the degree
of polarization of both electrons and neutrinos / anti-neutrinos. In the
case of electrons, the kinetic energy is found to be  very high; therefore,
the temperature of the system should be low enough to make them align. Since
the main mechanism of NS/PNS cooling is by the emission of neutrinos,
which in turn makes the neutrinos non-degenerate, and as a consequence the
condition as shown above in eqn.(10) will no longer be valid. Instead, one has 
to consider eqn.(6), for which there will be no spontaneous magnetization,
i.e., the process of alignment of electrons by reducing the temperature
of the system through neutrino emission destroys the possibility of
ferromagnetic transition.
Therefore, it is very difficult to have a ferromagnetic phase transition
even in PNS if the kinetic energy of the electronic part  is high enough,
which is sufficient to prevent spin alignment. If we concentrate on neutrino 
components, again there will
be more serious problem than the electronic sector, when we consider the 
spin alignment of neutrinos. In the case, when the decaying neutron has
spin up (down) and the recoiled proton has spin down (up), there will be
some directional constraint on the motion of trapped neutrino /
anti-neutrino (analogous to the famous $^{60}Co$ experiment) sum of electron 
and anti-neutrino spins should be $+1$ ($-1$).  There will also be left-handed 
to  right-handed  and vice-versa flipping problem, unless the neutrinos / 
anti-neutrinos carry some non-zero mass.

Therefore, we conclude that in a very old neutron star, whose
temperature is low enough and the neutrinos are non-degenerate, the
spontaneous ferromagnetic transition is absolutely forbidden. On the
other hand, in a PNS, where the neutrinos remain trapped within the
matter for quite a long duration, are degenerate, both the electronic
and neutrino components play major roles in ferromagnetic phase
transition. The electron kinetic energy should be extremely low for
their spin alignment, whereas, the neutrinos should have some finite
non-zero mass (no definite helicity state) to have left-handed to
right-handed spin flip and vice versa to make neutrinos / anti-neutrinos
also spin aligned. Therefore, the final conclusion is that the neutrino
mass has a significant role in spontaneous ferromagnetic phase
transition in PNS. There will be  no such transition if the neutrinos
are mass less.

On the other hand for NS matter with arbitrary ISO-spin asymmetry parameter
$x=(n_n-n_p)/n_B$, where $n_B=n_n+n_p$, the total baryon number density
of the system, there could be a spontaneous ferromagnetic transition
\cite{R10}, however, the system will again become
unpolarized as soon as $\beta$-equilibrium is reached, which we cannot
avoid as the NS matter in $\beta$-equilibrium is the only stable
configuration.

Therefore, if by some means a ferromagnetic transition occurs in PNS,
then in the latter stage of evolution, when neutrinos become
non-degenerate, there will be a spontaneous spin symmetry
restoration. We may therefore conclude that the origin of residual
magnetic field or the magnetic field of millisecond pulsars, which are
believed to be very old neutron stars ($\sim 10^8$G), is not 
ferromagnetic transition. Now it is quite possible to have a crust of magnetic
materials (e.g., iron) in such cold objects. To get an idea of what is
the contribution of such magnetic materials on the bulk magnetic field
of the object, we proceed in the following simplified manner. We give an
estimate on the upper limit of polar magnetic field strength. If $\vec
m$ is the total effective magnetic dipole moment of the crustal matter
due to tiny atomic magnets (which are assumed to be aligned), we have the 
field strength at the pole
\begin{equation}
B_p=\frac{2\mid \vec m\mid }{R^3}
\end{equation}
where $R$ is the radius of the star (assuming spherical). Assuming
maximal projection of the total magnetic moment in unit of $\mu_B$, the
Bohr magneton, we have
\begin{equation}
B_p=\frac{8\pi}{3}gJ\mu_Bn\left [ 1-\left (\frac{R'}{R}\right )^3 
\right ]
\end{equation}
where $R'$ is the thickness of the spherical crustal shell, $g$ is the
Lande $g$-factor and $J$ is the total angular momentum of the electron
and $n$ is the density of tiny magnets. Then for iron $n=\rho/M$ and
$gJ=6$, where $\rho$ is the density of crustal matter, which is assumed to be
$10^{10}$gm cm$^{-3}$ and $M$ is the mass of tiny atomic magnets. Putting
all these numerical numbers, we have seen that $B_p\approx 10$Tesla,
which is of course too low. Again, this is the upper limit of polar
magnetic field strength due to ferromagnetic transition of crustal
magnetic material. Therefore, the ferromagnetic transition in the
crustal matter can not produce magnetic field of the order
$10^7-10^8$G.

\noindent {\sl{Acknowledgment: SC is thankful to the Department of Science and 
Technology, Govt. of India, for partial support of this work, Sanction 
number:SP/S2/K3/97(PRU). SC is also acknowledges  Kamales Kar and Palas
Pal of SINP for some useful discussions and comments}}.  
\end{document}